\documentclass[aps,prd,preprint,groupedaddress]{revtex4-1}
\usepackage{amsmath}
\usepackage{amsfonts}
\usepackage{amssymb}
\usepackage{graphicx}

\begin{document}
\title{Gravitational waves from inflation with antisymmetric tensor field }

\author{Sandeep Aashish}
\email[]{sandeepa16@iiserb.ac.in}

\author{Abhilash Padhy}
\email[]{abhilash92@iiserb.ac.in}

\author{Sukanta Panda}
\email[]{sukanta@iiserb.ac.in}

\affiliation{Department of Physics, Indian Institute of Science Education and Research, Bhopal 462066, India}

\begin{abstract}
We build upon the past studies of inflation with rank-2 antisymmetric tensor field, including here the tensor perturbations to metric. We perform a comprehensive analysis of the background dynamics of our model in the presence of non-minimal coupling curvature terms $R$ and $R_{\mu \nu}$. We find appropriate conditions on the nonminimal coupling parameters to satisfy the constraint of speed of propagation of gravitational waves. Including the tensor perturbations, the model is found to be free from ghost instabilities with minimal constraints on the parameters. We also study the evolution of gravitational waves, calculate the power spectrum and the tensor spectral index.
\end{abstract}
\maketitle

\section{Introduction}
The inflationary paradigm introduced by Guth, rescues the standard bigbang model from observational inconsistency by providing reasonable explanation to the horizon and flatness problem \cite{guth1981}. There has been a lot of effort to build a model of inflation which meets the requirement of CMB observations. But the most recent high precision data of CMB  by Planck \cite{iacconi2019,akrami2019,akrami2018,contaldi2013} rules out most of the scalar field inflation models, especially single field inflation models \cite{linde1983,halliwell1986,gottlober1990,roberts1994,parsons1995,barrow1995,gomes2018,mazenko1986,kinney1995}. Moreover, the swampland criteria in string theory which sets some theoretical constraints for UV completion of any effective field theory, puts additional restriction on the scalar field potentials \cite{brennan2017, andriot2018, krishnan2018, obied2018, kallosh2019}. Alternative models of inflation based on vector field, face severe pathological issues like ghost and gradient instabilities \cite{ford1989, ryonamba2017, riotto2000, golovnev2008, darabi2014, bertolami2015}. Though there are few models that fit into the observational requirements but they are heavily constrained \cite{linde1993,sheikhjabbari2013}. Yet another set of theoretically sound inflation models free from the problems faced by vector inflation, are gauge-flation models constructed using non-abelian gauge fields \cite{jabbari2011,jabbari2012,jabbari2013a,jabbari2013b}. However, they have been shown to be in tension with the data \cite{peloso2013}. This has motivated attempts to build inflationary models with higher rank tensor fields, in particular 2- and 3-form fields. 2-forms appear naturally in superstring models \cite{rohm1986,ghezelbash2009} in low energy limits and are also referred to as the Kalb-Ramond field. Early studies of such n-form inflation models are carried out in Refs. \cite{koivisto2009,germani2009,prokopec2005,prokopec2006}, where 3-form field is found to be favorable over 2-form inflation due to the vector-inflation like ghost instabilities appearing in the later model.  More recently antisymmetric tensor has been studied in the context of $F(R)$ theories in Refs. \cite{elizalde2018, elizalde2019}.  
  
In contrast to the past conclusions, the results of recent studies \cite{aashish2018,aashish2019} have shown that slow roll inflation is indeed supported by rank-2 antisymmetric tensor field (2-form) when nonminimal coupling terms are included. Specifically, the presence of nonminimal coupling terms with the Ricci scalar $R$ and/or the Ricci tensor $R_{\mu \nu}$ is a sufficient condition for the existence of de-sitter solutions, thereby supporting inflation \cite{aashish2018}. Furthermore, known ghost and gradient instabilities can be avoided at least for the perturbed modes of $B_{\mu\nu}$ (keeping the metric unperturbed) by incorporating a gauge symmetry breaking kinetic term into the action \cite{aashish2019}.

In this paper we extend our study of the cosmological perturbation theory of 2-form inflation starting with the inclusion of tensor perturbations to metric, usually referred to as the primordial gravitational waves, in the model prescribed in \cite{aashish2019}. The availability of recent gravitational wave data coming from binary neutron star merger GW170817 and its associated electromagnetic counterparts, demands that the speed of propagation of gravitational waves be equal to the speed of light \cite{cai2016}. We study the constraints on the coupling parameters of our theory, for which this condition is satisfied, and find that unlike several inflation models \cite{kobayashi2009,mondal2020, guzzetti2016, odintsov2019, ito2016, obata2018}, this requirement is easily achieved by constraining one of the nonminimal coupling parameters. We would like to point out that there exists a class of inflation models for which the speed of gravitational waves can be made equal to unity (in natural units) through a set of conformal and disformal transformations of the metric \cite{creminelli2014n}. However, in our case this is achieved by constraining the parameters of our theory, not metric transformations. For completeness, we also address a past issue \cite{aashish2018} related to the parameter space for stable de-sitter solutions, and check its consistency with an instability analysis of perturbed modes including tensor perturbation. The analysis of scalar and vector perturbations is not included here, and will be addressed in a future work. 

The organization of the paper is as follows. In Sec. \ref{sec2}, we study the background cosmology of our model considering the contribution from both the nonminimal coupling term $R$ and $R_{\mu \nu}$. In Sec. \ref{sec3} we introduce tensor perturbation into the action and check the existence of ghost instability in the quadratic ordered part of the perturbed action. We also calculate the speed of propagation of the gravitational wave in this section and have given our prescription to make it unity. In Sec. \ref{sec4}, we solve the gravitational wave equation and have studied the behavior of gravitational waves in subhorizon and superhorizon limits. Along with it we calculate the tensor power spectrum and the tensor spectral index in this section. We conclude in Sec. \ref{concl} with some future prospects of this work.


\section{\label{sec2}Background Cosmology}
In this section we review and generalize the background analysis of Refs. \cite{aashish2018,aashish2019}. Previously it was shown that a stable slow roll inflation could be achieved with a rank-2 antisymmetric tensor field by simply including either of the nonminimal coupling with curvature terms $R$ and $R_{\mu \nu}$. Here, we generalize the previous background analysis  by including both $R$ and $R_{\mu \nu}$ couplings so that we achieve more freedom of the parameters. In the forthcoming section the need of this extra freedom will be evident when we try to match the speed of propagation of the gravitational wave with the observational expectation . The general form of the action for our model is given by
  \begin{eqnarray}
\label{bc1}
S = \int d^4x \sqrt{-g} \Big[\dfrac{R}{2\kappa} - \frac{1}{12} H_{\lambda \mu \nu} H^{\lambda \mu \nu} + \dfrac{\tau}{2} (\nabla_{\lambda} B^{\lambda \nu})(\nabla_{\mu} B^{\mu }_{\ \nu}) + (\frac{\xi}{2\kappa}R - \frac{m^2}{4}) B^{\mu \nu}B_{\mu \nu} \nonumber \\ + \frac{\zeta}{2\kappa} B^{\lambda \nu }B^{\mu}_{\ \nu} R_{\lambda \mu}  \Big].
\end{eqnarray}
   Where $B_{\mu \nu}$ is the antisymmetric tensor field and $H_{\lambda \mu \nu} = \nabla_{\lambda} B_{\mu \nu} + \nabla_{\mu} B_{\nu \lambda} + \nabla_{\nu} B_{\lambda \mu }$  is  the field strength . '$g$' symbolizes the determinant of the metric $g_{\mu \nu}$ and $\kappa $ denotes the square inverse of the reduced Planck's mass $M_{Pl}$. The first and second term in the action (\ref{bc1}) express respectively the gauge invariant kinetic term and  the gauge symmetry violating kinetic term that rescues the system from the ghost and gradient instability. Rest of the terms in the action (\ref{bc1}) include the nonminimal coupling with the curvature terms $R$ , $R_{\mu \nu}$ and the self interacting quadratic potential term.
                                     Assuming  the early universe to be isotropic and homogeneous , the background metric  is chosen to be the FLRW metric with components
 \begin{equation}
 \label{bc2}
 g_{00} = -1,  \quad  g_{ij} = a(t)^2 \delta_{ij},
 \end{equation}
  and the background antisymmetric tensor field $B_{\mu \nu}$ is structured as
 \begin{equation}
 \label{bc3}
  B_{\mu \nu} =  a(t)^2 \phi(t)\begin{pmatrix}
 0 & 0 & 0 & 0 \\
 0 & 0 & 1 & -1 \\
 0 & -1 & 0 & 1 \\
 0 & 1 & -1 & 0
 \end{pmatrix}.
 \end{equation}
 \begin{figure}

\includegraphics[scale=0.5]{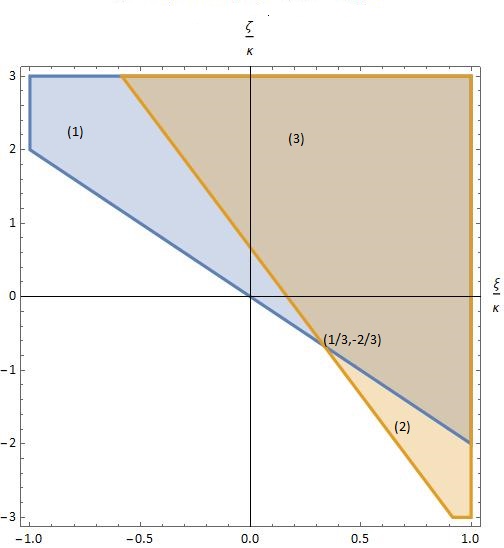}
\caption{\label{fig1} This  plot  shows the allowed  values of ($\xi , \zeta $)  for the existence of exact de-Sitter solution in the model. Regions 1 and 2 exhibit the portion in the parameter space where   $3\zeta + 6\xi > 0$ and $ 3\zeta + 12\xi > 2 \kappa $  are individually satisfied  respectively  where as  region 3 exhibits the portion where both the conditions $3\zeta + 6\xi > 0$ and $ 3\zeta + 12\xi > 2 \kappa $  hold true.}
\end{figure}
  The time dependence of the scale factor $a(t)$, the Hubble parameter $H(t)$ and the field $\phi(t)$ is to be assumed hereafter until and unless specified otherwise.  With this choice of $g_{\mu \nu}$ and $B_{\mu \nu}$,  the equations of motion appears similar to  the case of scalar field inflation model.  It should be noted that the gauge breaking kinetic term in the action (\ref{bc1}) does not have any effect on the background dynamics \cite{aashish2019}. The dynamics of this system is described by the Einstein's equation $G_{\mu \nu} = \kappa T_{\mu \nu}$ and the $B_{\mu \nu}$ field equation along with a constraint equation for conservation of energy momentum tensor($\nabla^{\mu} T{\mu \nu} = 0 $). Using the constraint equations and some manipulations, one obtains two independent equations \citep{aashish2018} which are as follows:
 \begin{eqnarray}
\label{bc4}
H^2  = \frac{\kappa}{2}[(\dot{\phi} + 2 H\phi)^2 +m^{2}\phi^{2}]
-6\xi (2H\phi \dot{\phi} + H^2 \phi^2)  - 2\zeta H\phi \dot{\phi},\\ 
\label{bc5}
2\dot{H} + 3H^2 + \left(12\xi +\frac{3}{2} \zeta \right) (\phi \ddot{\phi} + \dot{\phi}^2) +  \left(24\xi -\frac{3}{2} \zeta \right)H\phi \dot{\phi}
 -  (6\xi +3 \zeta) \dot{H}\phi^2 \nonumber\\
 -  (18\xi +9 \zeta) H^2 \phi^2 = 0.
\end{eqnarray}
 An exact  de-Sitter type inflation where $H$ and $\phi $ both remain constant could be realized by demanding certain conditions over the coupling parameters $\xi$ and $\zeta$. Those conditions in various cases are listed in Table \ref{tab1}. These constraints are also graphically shown by the shaded region  over  the plane of $(\xi / \kappa , \zeta / \kappa)$  in the Fig. \ref{fig1}. It can be observed that for $\xi \leq \frac{\kappa}{3}$ the condition $12 \xi + 3\zeta > 2\kappa$ is sufficient for getting de-Sitter solution and similarly for $\xi > \frac{\kappa}{3}$, the condition $2\xi + \zeta > 0$ is sufficient .

 \begin{table}[h!]
  \begin{center}
    \begin{tabular}{|c|c|c|c|} 
    \hline
     Cases & \ $\phi_{0}^{2}$ \ & $H_{0}^{2}$ & Condition\\
      \hline
     $\xi \neq 0 ,\  \zeta = 0$ & $\dfrac{1}{6\xi}$ & $\dfrac{\kappa m^{2}}{4(6\xi - \kappa)}$ & $\xi > \dfrac{\kappa}{6}$\\
      \hline
     $\xi = 0,\ \zeta \neq 0 $ & $\dfrac{1}{3\zeta}$ & $\dfrac{\kappa m^2}{2(3 \zeta - 2 \kappa)}$ & $\zeta > \dfrac{2 \kappa}{3}$\\
      \hline
     $\xi  \neq  0,\ \zeta \neq 0 $ & $\dfrac{1}{3\zeta + 6\xi}$ & $\dfrac{\kappa m^2}{2(3 \zeta + 12\xi  - 2 \kappa)}$ & $3\zeta + 6\xi > 0,\ 3\zeta + 12\xi > 2 \kappa $\\
      \hline
    \end{tabular}
    \caption{The de-Sitter space solutions of $\phi$ and $H$, along with the condition on parameters $\xi$ and $\zeta$ corresponding to $R$ and $R_{\mu\nu}$ coupling terms respectively.}
    \label{tab1}
  \end{center}
\end{table}
 
\subsection{Stability of the de-Sitter background}
 Ideally one can not expect a steady exact de-Sitter type background rather it  may evolve to a quasi de-Sitter type scenario where the Hubble parameter $H$ and $\phi$ have little fluctuations $\delta H$ and $\delta \phi $ instead of remaining constant permanently. But for a stable de-Sitter background, these fluctuations  should not diverge with time but are expected to die out after a while. The behavior of these fluctuations can be analyzed by solving the system of dynamical equations
  \begin{eqnarray}
  \label{bc6}
 \dfrac{d}{dt} 
   \begin{pmatrix}
   \delta \phi \\
   \delta H 
   \end{pmatrix}  = A \begin{pmatrix}
   \delta \phi \\
   \delta H 
   \end{pmatrix} ,
  \end{eqnarray}
    Where $A$ is a $(2 \times 2)$ square matrix, whose components are expressed as
    \begin{eqnarray}
    \label{bc7}
   & A_{11} = \left(\dfrac{6\xi + 3\zeta}{6\xi + \zeta - \kappa}\right) H_0,\quad 
      A_{12} = - \left(\dfrac{12\xi + 3\zeta-2\kappa}{6\xi + \zeta - \kappa}\right) \phi_0,\nonumber\\\\
         &A_{21} = -\dfrac{1}{(6\xi + \zeta - \kappa)}\left(\dfrac{(6\xi + 3\zeta)(8\xi + \zeta )+(6\xi + \zeta - \kappa)(16\xi - \zeta )-4(6\xi + \zeta - \kappa)^2}{(6\xi + \zeta - \kappa)12\xi -\zeta (8\xi + \zeta )}\right)\dfrac{H_0^2}{\phi_0^3} ,\nonumber\\\\
         & A_{22} =  \dfrac{(12\xi+3\zeta - 2\kappa)}{(6\xi + \zeta - \kappa)}\left(\dfrac{(6\xi + \zeta)(8\xi + \zeta )+(12\xi +3 \zeta -2 \kappa)(16\xi - \zeta )}{(6\xi + \zeta - \kappa)12\xi -\zeta (8\xi + \zeta )}\right)H_0.
   \end{eqnarray}
   
       \begin{figure}
\includegraphics[scale=0.7 ]{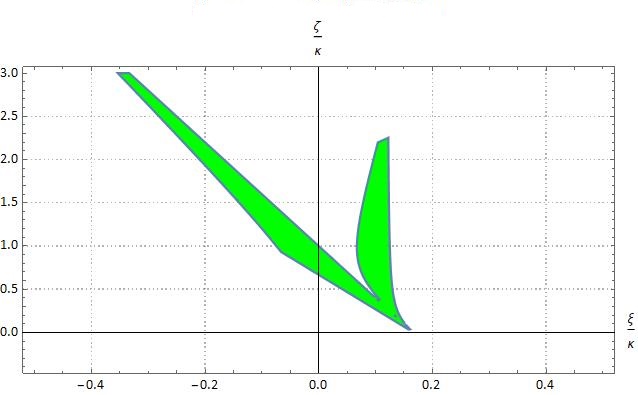},
\caption{\label{fig2} The V shaped region in the plot indicates the allowed region in the plane of parameters $(\xi / \kappa , \zeta / \kappa)$ , where the real part of both the eigen values $\lambda_1$ and $\lambda_2$ are negative and the conditions for exact de-Sitter solutions hold true.}
\end{figure}
The system of equations presented in Eq. (\ref{bc6})  are obtained from the background equation of motion Eq. (\ref{bc4}) and Eq. (\ref{bc5}), where we substitute $H = H_0 + \delta H$ , $\phi = \phi_0 + \delta \phi$ and consider only the linear part of the equation for $(\delta \phi , \delta H)$. It is in the form of an Eigenvalue equation and the general solution to it can be obtained in the form  $ C_1 e^{\lambda_1 t}+C_2 e^{\lambda_2 t}$ where  $\lambda_1 , \lambda_2$ are the two eigen values of matrix $A$ . If real part of both the eigen values are negative, then $(H,\phi)$ eventually becomes  $(H_0, \phi_0)$ with time and the de-Sitter space becomes  stable . However this situation is not possible for any arbitrary $\xi$ and $\zeta $ rather it further constrains the parameter space of $\xi$ and $\zeta$, which is pictorially shown in Fig. \ref{fig2}.  So in this range of $\xi $ and $\zeta $ , the de-Sitter solutions  $(H_0,\phi_0 )$ behaves as stable fixed point.

For consistency, we also check the null energy condition (NEC) for our model, which requires,
    \begin{eqnarray}
    \label{bc8}
    T_{\mu \nu} n^{\mu} n^{\nu} > 0,
    \end{eqnarray}
 where $n^{\mu} , n^{\nu}$ are two null vectors. In our model, the energy-momentum tensor is given by \cite{aashish2018,aashish2019}
 \begin{eqnarray}
 \label{bc9}
T_{\mu \nu} = \dfrac{1}{2}H^{\alpha \beta}_{\quad \mu} H_{\nu \alpha \beta}+ m^2 B^{\alpha}_{\ \mu} B_{\alpha \nu} - g_{\mu \nu} (\frac{1}{12} H_{\alpha \beta \gamma} H^{\alpha \beta \gamma} + \frac{1}{4}m^2 B_{\alpha \beta}B^{\alpha \beta})\nonumber\\
+\dfrac{\xi}{\kappa}\left[\nabla_{\mu} \nabla_{\nu}(B_{\alpha \beta}B^{\alpha \beta}) - g_{\mu \nu} 
\nabla^{\lambda} \nabla_{\lambda} (B_{\alpha \beta}B^{\alpha \beta}) - G_{\mu \nu} (B_{\alpha \beta}B^{\alpha \beta}) - 2R  B^{\alpha}_{\ \mu} B_{\alpha \nu}   \right]\nonumber\\
+ \frac{\zeta}{\kappa} \Big[ \frac{1}{2} g_{\mu \nu} (B^{\alpha \gamma} B^{\beta}_{\ \gamma}R_{\alpha \beta} - \nabla_{\alpha} \nabla_{\beta} B^{\alpha \gamma} B^{\beta}_{\ \gamma}) - B^{\alpha }_{\ \mu} B^{\beta}_{\ \nu}R_{\alpha \beta}
  -B^{\alpha \beta} B_{\mu \beta}R_{ \nu \alpha} - B^{\alpha \beta} B_{\nu \beta}R_{ \mu \alpha}&& \nonumber \\ 
 + \frac{1}{2}(\nabla_{\alpha} \nabla_{\mu} B_{\nu \beta} B^{\alpha \beta} +\nabla_{\alpha} \nabla_{\nu} B_{\mu \beta} B^{\alpha \beta} - \nabla^{\lambda} \nabla_{\lambda} B^{\alpha}_{\ \mu} B_{\alpha \nu})  \Big]\nonumber\\
 +  \dfrac{\tau}{2} \Big[ g_{\mu \nu} \left(  (\nabla_{\lambda} B^{ \sigma \lambda })(\nabla_{\rho} B^{\rho }_{\ \sigma}) + 2 B^{\sigma \lambda} \nabla_{\lambda} \nabla_{\rho} B^{\rho}_{\ \sigma} \right) + 2 (\nabla_{\lambda} B^{\lambda}_{\ \mu})(\nabla_{\rho} B^{\rho}_{ \ \nu})\nonumber\\
     + 2\left( B_{\mu}^{\ \lambda} \nabla_{\lambda} \nabla_{\rho} B_{\nu}^{\ \rho} +  B_{\nu}^{\ \lambda} \nabla_{\lambda} \nabla_{\rho} B_{\mu}^{\ \rho} \right)  \Big].
\end{eqnarray}
The choice of null vector is governed by the property $g_{\mu \nu} n^{\mu} n^{\nu} = 0$, following which we choose,
\begin{eqnarray}
  \label{null}
  n^{\mu} = \left(1 , \frac{1}{\sqrt{3}a}, \frac{1}{\sqrt{3}a}, \frac{1}{\sqrt{3}a}\right).
\end{eqnarray}
Using Eqs. (\ref{null}) and (\ref{bc9}) we obtain,
\begin{eqnarray}
\label{bc10}
T_{\mu \nu} n^{\mu} n^{\nu} &=& 3(1+ \frac{4\xi}{\kappa}) \dot{\phi}^2 + 12(1 - \frac{\xi + \zeta}{\kappa}) H\phi \dot{\phi} + 12 H^2 \phi^2 + \frac{12 \xi}{\kappa} (\phi \ddot{\phi} + \dot{H} \phi^2)\nonumber\\
&=&\left( 3 (1 + \frac{8\xi}{\kappa}) \delta^2 + \frac{12\xi}{\kappa H} \dot{\delta} + 12 (1 - \frac{\xi + \zeta}{\kappa})\delta - \frac{12 \xi}{\kappa} \epsilon + 12\right) H^2 \phi^2 \nonumber\\
& \approx & 12 (1 + \delta -  \frac{ \xi}{\kappa} \epsilon  - \frac{\xi + \zeta}{\kappa}\delta  )H^2 \phi^2.
\end{eqnarray}
Therefore the null energy condition is satisfied when
\begin{eqnarray}
\label{bc11}
\left[\frac{\xi + \zeta}{\kappa} - 1  \right] \delta + \frac{\xi}{\kappa} \epsilon < 1.
\end{eqnarray}
Clearly the null energy condition for our model depends on both slow roll parameters $\epsilon$ and $\delta$ (defined below in Eq. (\ref{tpa5})), which is in contrast to past results for scalar models, for example in Ref. \cite{creminelli2014n} where the NEC depends on $\epsilon$ alone. 

\section{\label{sec3}Tensor Perturbation}
A preliminary study of perturbations for this model was undertaken in Ref. \cite{aashish2019}, where all the calculations were done in a test frame in which the metric perturbations were ignored. In that setup, the theory is free from both the ghost and the gradient instability under the conditions that coupling parameter $\tau $ be positive and satisfies the following:
   \begin{eqnarray}
   \label{tp1}
   \tau > \dfrac{a^2 m^2}{k^2}  ,\quad
   \tau > -\left( 1 +  \dfrac{a^2 m^2}{2k^2}  \right) + \sqrt{ \dfrac{2a^2 m^2}{k^2} \left( 1+  \dfrac{a^2 m^2}{8k^2} \right)}.
   \end{eqnarray}

Though this preliminary investigation of instability in perturbed modes enhances the viability of the model but the more realistic approach will be to perform a  complete perturbation analysis where the metric perturbations are also included. From the metric side we get  four  scalar, four vector and two tensor modes of perturbation  where as   from the $B_{\mu \nu}$ side we get two scalar and four vector modes of perturbation. From the SVT decomposition \cite{riotto2000} , the general form of the metric perturbation is expressed as
   \begin{eqnarray}
   \label{tp2}
  &  g_{00} = -(1+ \psi)  \quad  \quad  g_{0i} = a(\partial_i \chi + E_i)\nonumber\\  
    & g_{ij} = a^2 \left[ (1-2\alpha) \delta_{ij} + 2 \partial_{ij} \beta + ( \partial_i F_j + \partial_j F_i ) + h_{ij} \right],
   \end{eqnarray}
   where $\psi , \chi , \alpha , \beta $ are the scalar modes, $E_i , F_i$ are the divergence free vector modes  and $h_{ij}$ is a traceless ($h_{ii} = 0$) and divergence free ($\partial_i h_{ij} = \partial_j h_{ij} = 0 $) matrix of order ($3 \times 3$). According to the decomposition theorem for cosmological perturbations the scalar, vector and tensor modes evolve independent of each other at the linearized level  so can be studied separately. As $B_{\mu \nu}$ does not contribute to the tensor perturbation it appears relatively simpler to study the tensor perturbation. Further any theory has to be independent of gauge transformations; so the gauge redundancy should be removed. The tensor mode  $h_{ij}$ is gauge invariant by default and one need not worry about the gauge redundancy. The tensor modes of perturbation $h_{ij}$ are usually referred to as primordial gravitational wave which carries the signature of the early universe. In this work, we study only the tensor perturbations leaving the vector and scalar perturbations to be studied in future. So the scalar and vector perturbations are ignored i.e $\delta g_{00} = 0 $ , $\delta g_{0i} = 0$ and $\delta g_{ij} = a^2 h_{ij}$. We can consider a simple structure for $h_{ij}$ as 
   \begin{eqnarray}
   \label{tp3}
   h_{ij}=
   \begin{pmatrix}
   h_+ & h_{\times} & 0\\
   h_{\times} & - h_+ & 0\\
   0 & 0 &0
   \end{pmatrix},
   \end{eqnarray}
    where the tracefree  and  transverse nature of $h_{ij}$ can be easily observed. Here the frame is  oriented in such a manner that the perturbation variables $h_{+}$ or $h_{\times}$ lies on the $(x,y)$ plane and the wave vector $\vec{k}$ is directed along $z-$ axis. We  express these two modes with a common symbol $h_e $ where  e can be  $+$ or $\times$.

\subsection{ Ghost Instability}
 Now we substitute the perturbed metric and the background structure of $B_{\mu \nu}$ as in Eq. (\ref{bc3}), in the action (\ref{bc1}). The action can be expanded upto second order in terms of the perturbation variables and  the second order part of it can be explicitly  written as
 \begin{eqnarray}
 \label{tpa1}
 S_2 = \int dt\  d^3x \  a^3 \left( 1- \dfrac{h_+^2 + h_{\times}^2}{2} \right)\Big[ \dfrac{R }{2\kappa} + \left( \dfrac{ \xi R }{6\kappa} - \dfrac{m^2}{12}  \right)\dfrac{(3+ 2 h_{\times} + 3(h_+^2 + h_{\times}^2))}{a^4} B_{ij}B_{ij} \nonumber\\
 - \dfrac{\zeta}{2\kappa} \dfrac{R_{ij}}{a^6} ( B_{ik}B_{kj} + Z_{ij} - W_{ij}) + \dfrac{\tau}{2a^6} B_{il} B_{jm} \partial_j h_{km} \partial_i h_{kl} + \nonumber\\
 + \dfrac{\dot{B}_{ij}^2}{a^4}\left( \frac{3}{2} + h_{\times}+\frac{3}{2} h_e^2 \right)\Big]
 \end{eqnarray}
  where $Z,W$ are matrices with components   $Z_{ij} = (BhB + B^2 h + hB^2 )_{ij}$ and $W_{ij} = (Bh^2 B + B^2 h^2 +h^2B^2+BhBh+ hB^2h + hBhB)_{ij}$. $Tr$ stands for trace of the matrix. To keep the  expression for action simple we don't write  $R$ and $R_{\mu \nu}$ explicitly. Notice that the kinetic term with  $\tau $ coupling only contributes to the spatial derivatives of $h_{ij}$, thereby avoiding additional constraints on $\tau$. The action \ref{tpa1} can be  Fourier transformed into momentum space, where all the modes with different momenta $k^i$  evolve independently. The structure of $h_{ij}$  is set for  the coordinate frame where the momentum  $k^i = (0,0,k)$. That takes away the intermixing of the $h_{+}$ and ${h_{\times}}$ modes in the action and the action is obtained as
       \begin{eqnarray}
       \label{tpa2}
       S_{2} = \sum_{e= + ,  \times }  \int dt d^3k   \frac{a^3}{4\kappa} \left[ \Omega_k \  \dot{h}_e^{\dagger} \dot{h}_e + \Omega_c \ (\dot{h}_e^{\dagger} h_e + h_e^{\dagger} \dot{h}_e) + \Omega_g \ h_e^{\dagger} h_e \right],
       \end{eqnarray}
  
\begin{figure}
\includegraphics[scale=0.7 ]{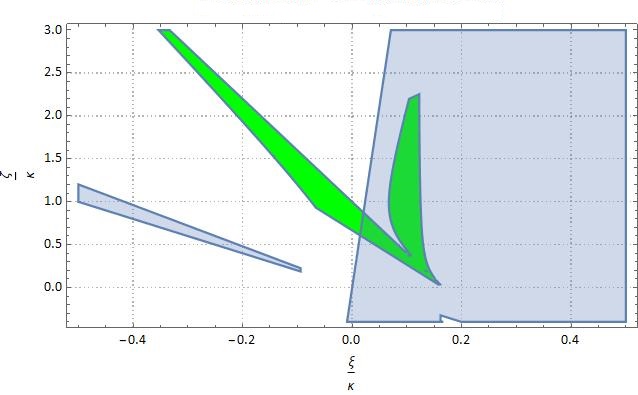},
\caption{\label{fig3} The shaded region in the graph represents the parameter space within which gravitational modes are free from instability and the V shaped region indicates the stable de-Sitter solution zone. We have taken  $\delta = 1 $ and $\epsilon = 1 $.  }
\end{figure}

     For simplicity in notations, the Fourier transform of $h_{(+/\times)}$ are also expressed with the same symbol  $h_{(+/\times)}$. It can be noticed that, both the decoupled part in action (\ref{tpa2}) for  $h_+$ and $h_{\times}$   take the same form  and same set of  coefficients $\Omega_k , \Omega_c , \Omega_g $ which are given by 
     \begin{eqnarray}
     \label{tpa3}
    & \Omega_k = \left[1 + 2(3\xi + \zeta) \phi^2 \right],\nonumber\\
   &  \Omega_c =  \left[ 6 (4\xi+\zeta) \phi \dot{\phi} -6(  2\xi + \zeta)H \phi^2 - 2H \right],\nonumber\\
     &\Omega_g = \Big[ 6(\zeta + 6 \xi) \dot{H} \phi^2 + 6(3\zeta + 12 \xi ) H^2 \phi^2 + 3 \kappa\left((\dot{\phi}+2H\phi)^2 - m^2 \phi^2\right)\nonumber\\
     &  - \frac{k^2}{a^2} \left( 1 + (6 \xi + 4\zeta - 4 \kappa \tau) \phi^2  \right) - 6(\dot{H} + 2 H^2 )\Big].
     \end{eqnarray}
        If there is no ghost  present in the theory, then the coefficient of the kinetic energy term $\Omega_k$ needs to be positive always. In the de-Sitter limit $\phi^2 $ receives a constant value of  $\phi_0^2 = 1/(6\xi + 3\zeta)$ so that $\Omega_k = (12\xi + 5\zeta)/(6\xi+3\zeta) $.
And we escape the problem of ghost instability by simply demanding $(12\xi + 5 \zeta) > 0$. However in the quasi-de-Sitter scenario, $\phi^2 $ is not exactly a constant but varies slightly from the de-Sitter value and can be expressed in terms of slow roll parameters as      
 \begin{eqnarray}
 \label{tpa4}
 \phi^2 = \dfrac{1}{(6\xi + 3\zeta)}\left[ 1 - \dfrac{\epsilon}{2}+ \dfrac{(16\xi - \zeta)}{(6\xi + 3\zeta)} \dfrac{\delta}{2} \right],
 \end{eqnarray}
 where $\epsilon$ and $\delta $ are the two  slow roll parameters and are given by
       \begin{eqnarray}
       \label{tpa5}
       \epsilon = - \frac{\dot{H}}{H^2}, \quad \delta = \frac{\dot{\phi}}{H\phi}.
       \end{eqnarray}
       This expression for $\phi^2 $ in Eq. (\ref{tpa4}), is obtained from  Eq. (\ref{bc4}) and Eq. (\ref{bc5}) where we neglect  the $\ddot{\phi}$ and $\dot{\phi}^2$ term along with the nonlinear terms of slow roll parameter. As the field is rolling slowly over the potential, this approximation is valid.  In this approximation we obtain
       \begin{eqnarray}
       \label{tpa6}
       \Omega_k =  \dfrac{(12\xi + 5\zeta)}{(6\xi + 3\zeta)} - \dfrac{(3\xi + \zeta)}{(6\xi + 3\zeta)} \epsilon +\dfrac{(16\xi -\zeta)(3\xi + \zeta)}{(6\xi + 3\zeta)^2} \delta.
       \end{eqnarray}
       As the value of $ \epsilon $ varies between 0 to 1, in order to keep $\Omega_k $ always positive we need to have
       \begin{eqnarray}
       \label{tpa7}
       \dfrac{12\xi + 5\zeta}{3\xi + \zeta} + \dfrac{16\xi - \zeta}{6\xi + 3\zeta}  \ \delta \  > \  1,
       \end{eqnarray}
      where we have taken the value of $\epsilon $ to be its maximum i.e unity. In Fig. \ref{fig3}, it is shown that in the first quadrant of the $(\xi , \zeta)$ parameter plane, the constraints for  stable de-Sitter solution fits well into the requirement for removing ghost instability. 
      
\subsection{Speed of the Gravitational waves} 
Though we can evade from the ghost instability by setting some constraints on the coupling parameters, still there can be pathological instabilities in the gravitational wave if the square of the speed of propagation of the gravitational waves turns negative. The objective of this section is twofold: first, we check the basic requirement that the squared speed of the gravitational wave must be real positive, and second, we obtain constraints so that the gravitational wave speed becomes unity in accordance with the recent gravitational wave data GW170817  \cite{odintsov2019}.  
The equations of motion for the gravitational waves can be obtained by  varying the action (\ref{tpa2}) with respect to $h_e^{\dagger}$
      \begin{eqnarray}
      \label{tpb1}
      \ddot{h}_e + \left( \dfrac{\dot{\Omega_k}}{\Omega_k} + 3H \right)\dot{h}_e
+ \left( \dfrac{\dot{\Omega_c}+ 3H \Omega_c - \Omega_g}{\Omega_k} \right) h_e = 0, \ (e = +/\times) .    
 \end{eqnarray}
Eq. (\ref{tpb1}) looks like a damped harmonic oscillator except the coefficients of $\dot{h_e}$ and $h_e$ are not constants. These coefficients are dependent on time, and the time dependence is coming from the terms $\phi^2 $ and $H^2$. In quasi de-Sitter space they can be expressed in terms of the slow roll parameters $\epsilon$ and $\delta$ and can be approximated to be constant. The expression for $\phi^2 $ and $H^2$ in quasi de-Sitter space is given by 
     \begin{eqnarray}
       \label{tpb2}
   &\dfrac{1}{\phi^2} \approx (\zeta + 2\xi)(3+\epsilon) + \left(\dfrac{\zeta}{2} -8\xi \right) \delta, \nonumber\\
    &  \dfrac{\kappa m^2}{H^2} \approx 2(12\xi + 3\zeta - 2\kappa) + 2(\zeta + 2\xi) \epsilon + (8\xi + 5\zeta - 2\kappa)\delta.
       \end{eqnarray}
       Now substituting Eq. (\ref{tpb2}) in Eq. (\ref{tpb1}), the coefficients can be re expressed as 
  \begin{eqnarray}
  \label{tpb3}
  \dfrac{\dot{\Omega_k}}{\Omega_k}  =  \dfrac{(12\xi + 4\zeta) \delta}{(12\xi + 5\zeta)}H, \quad 
 \dfrac{\dot{\Omega_c}+ 3H \Omega_c - \Omega_g}{\Omega_k} =  \dfrac{k^2 F}{a^2},
  \end{eqnarray}
Where  $F$  is written as 
 \begin{eqnarray}
 \label{tpb5} 
  F = 1 +  \dfrac{( \zeta - 2\kappa \tau )}{(12\xi + 5 \zeta)^2} \left[2(12\xi + 5\zeta)-2(\zeta + 2\xi)\epsilon - (\zeta - 16 \xi)\delta  \right].
  \end{eqnarray}
The dispersion relation can be derived by substituting a solution of the form
$h_e \propto \exp [-i \int^t ( c_T k /a(t')) dt']\vec{e}$ in the equation of motion Eq. (\ref{tpb1}). $c_T$ is the speed of the gravitational wave  and $\vec{e}$ is a constant vector.  The dispersion relation can be expressed as a quadratic equation in terms of $c_T $ as 
   \begin{eqnarray}
   \label{tpb6}
   c_T^2 + i \left( \frac{2aH}{k} \right)\left( 1 + \dfrac{6\xi + 2\zeta}{12\xi + 5\zeta} \delta\right) c_T - F = 0.
   \end{eqnarray}
  
 In the deep subhorizon limit i.e when $k >> aH$, the second term in Eq. (\ref{tpb6}) can be neglected  and $c_T$ can be expressed as 
   \begin{eqnarray}
   \label{tpb7}
  & c_T^2 = F  =  1 + \dfrac{2(\zeta - 2 \kappa \tau)}{(12\xi + 5 \zeta)} \left( 1 -  \dfrac{(\zeta + 2 \xi)}{(12\xi + 5 \zeta)} \epsilon +  \dfrac{(16 \xi -\zeta )}{2(12\xi + 5 \zeta)} \delta \right).
   \end{eqnarray}
    The amplitude of the primordial GWs is determined by $c_T$ and the Hubble radius $\sim H^{-1}$ which can have strong impact on the power spectrum \cite{cai2016}. The recent observation from neutron star merger data GW170817 \cite{odintsov2019}  insists the gravitational wave speed to be equal to the speed of light i.e unity in natural units. A number of cosmologically viable theories predicting variable  gravitational wave speed, in particular covariant Gallileon and their Horndeski extensions as well as some other modified gravity theories, do not fulfill this requirement and are disfavored  \cite{ezquiaga2017}. In a study of p-form inflation, the speed of the gravitational wave differs from unity  for 2-form  when the mass of the field is tachyonic \cite{kobayashi2009}. But in our case it can be noticed  from the expression of $c_T^2$ in Eq. (\ref{tpb7}) that $c_T$ can be approximated to unity if  $\zeta \approx 2 \kappa \tau$ . So at this stage it becomes important to check whether this demand does not contradict the previous constraints on $\tau $ and $\zeta $. Using the conditions on $\tau $ given in  Eq. (\ref{tp1}), the corresponding constraints on $\zeta $ are
    \begin{eqnarray}
    \label{tpb8}
   \zeta > 2\kappa \dfrac{a^2 m^2}{k^2} = 2\kappa \dfrac{a^2 H^2}{k^2}  \dfrac{m^2}{H^2} ,\nonumber\\
   \zeta > 2\kappa \left[ -\left( 1 +  \dfrac{a^2 m^2}{2k^2}  \right) + \sqrt{ \dfrac{2a^2 m^2}{k^2} \left( 1+  \dfrac{a^2 m^2}{8k^2} \right)}\right],
\end{eqnarray}  
in order to get $c_T^2 = 1$.
 In subhorizon limit and the superhorizon limit the second condition on $\zeta$ in Eq. (\ref{tpb8}) can be rewritten as $\zeta > 2 \sqrt{2} \kappa (am/k)$ and $\zeta > 2\kappa $ respectively while the first condition demands a very large value of $\zeta $ in superhorizon limit. But from Fig. \ref{fig3}, it is observed that the  allowed values of $\zeta $ have a finite range $[0.1 \kappa , \  2.3 \kappa]$. This requirement can be fulfilled if, for example, mass of the field $m$ is small in comparison to the Hubble parameter. From Eq. (\ref{tpb2}), it can be noticed that if we assume $m < H $ then we get the condition 
\begin{eqnarray}
\label{tpb9}
\left( \dfrac{12\xi + 3 \zeta}{\kappa} - 2 \right) + \left( \dfrac{\zeta + 2\xi}{\kappa} \right) \epsilon + \left( \dfrac{8\xi + 5\zeta}{2\kappa} - 1 \right) \delta < \dfrac{1}{2}.
\end{eqnarray}
As  $\epsilon $ and $\delta $ take very small values, it satisfies the condition $12\xi + 3\zeta < 2.5 \kappa$. This condition does not contradict the previous constraints of $\zeta $ rather it makes the constraint more strict, i.e. $ 2\kappa < 12 \xi + 3\zeta < 2.5 \kappa$. 
The importance of  the nonminimal coupling terms $B^{\lambda \nu }B^{\mu}_{\ \nu} R_{\lambda \mu}$ in the action can be realized here. In absence of this term, it would not have been possible to achieve $c_T = 1$.  Here onwards we will take $c_T = 1$ in our calculation.

\section{\label{sec4}Evolution of  the gravitational wave equation}
In the previous section we showed  that gravitational waves are free from instabilities. Now  we can find the solutions of the gravitational wave equation  Eq. (\ref{tpb1}) and can analyze it's behavior at different stages of inflation. It is useful to express Eq. (\ref{tpb1}) in terms of conformal time coordinate $\eta$   \cite{riotto2000},  which is defined as 
\begin{eqnarray}
\label{ega1}
\eta \equiv \int_{t_e}^t  \dfrac{dt'}{a(t')},
\end{eqnarray}      
where $t_e$ is denotes the time at which inflation ends and $t$ is any arbitrary time. As a result of this definition  $\eta$  is  negative during inflation. The modes cross the horizon at   $k |\eta | =  1$  while  $k |\eta | >  1$ and  $k |\eta | <  1$ describe subhorizon and superhorizon modes respectively. With this replacement of time coordinate Eq. (\ref{tpb1}) can be rewritten as
         \begin{eqnarray}
    \label{ega2}
    h_e'' + \left( 2 + \dfrac{12 \xi + 4\zeta}{12 \xi + 5 \zeta} \delta \right) aH h_e' +  k^2 c_T^2 h_e = 0,
    \end{eqnarray}
    where the prime over $h_e$ represents derivative with respect to $\eta$ . We redefine $h_e$ as
    \begin{eqnarray}
    \label{ega3}
     h_e \to \tilde{h}_e = a^{\lambda} h_e, \quad  \lambda =  1+ \dfrac{6 \xi + 2\zeta}{12 \xi + 5 \zeta}\delta ,
    \end{eqnarray}
    where $a$ is the scale factor.
    Then the equation becomes
    \begin{eqnarray}
    \label{ega4 }
    \tilde{h}_e'' + \left[ k^2 +(1+\epsilon - 3\lambda) a^2 H^2\right] \tilde{h}_e = 0,
    \end{eqnarray}
   where  we take $c_T^2 = 1$.  In quasi de-Sitter space,  $aH \approx - (1+ \epsilon)/\eta$. So Eq. (\ref{ega3}) becomes
    \begin{eqnarray}
    \label{ega5}
        \tilde{h}_e'' + \left[ k^2 - \dfrac{\omega^2 }{\eta^2}\right] \tilde{h}_e = 0,\quad  \omega^2 = 3(\lambda + \epsilon ) - 1.
    \end{eqnarray}
      
    Eq. (\ref{ega5}) resembles  a quantum harmonic oscillator equation. So  $\tilde{h}_e$  can be written  in terms of the creation and annihilation operator as
    \begin{eqnarray}
    \label{ega6}
    \hat{\tilde{h}}_e(k, \eta) = v_e(k,\eta) \hat{a}_{\vec{k}} + v_e^{\ast}(k,\eta) \hat{a}^{\dagger}_{\vec{k}},
    \end{eqnarray}
   where $\hat{a}_{\vec{k}}$ and $\hat{a}^{\dagger}_{\vec{k}}$ are the annihilation and creation operator for a mode with wave number $k$ respectively. The coefficients  $v_e(k, \eta)$ satisfies the equation 
    \begin{eqnarray}
    \label{ega7}
    v_e'' + \left(  k^2 - \frac{\omega^2}{\eta^2}\right) v_e = 0.
    \end{eqnarray}
  The initial condition to solve the problem is obtained from the natural hypothesis, where we assume that the Universe was in the vacuum state deﬁned as $\hat{a}_{\vec{k}} |0\rangle = 0$ \ at  very early stage, that is the “Bunch-Davies vacuum state” \cite{baumann2009}. When we observe the nature of the solutions of Eq. (\ref{ega7}) separately during the subhorizon and superhorizon regime, we find the second term in Eq. (\ref{ega7}) dominates over the third term  in  subhorizon limit and the solution is an oscillatory solution.  i.e 
   \begin{eqnarray}
   \label{ega8}
v_e = A e^{- i  k\eta},
  \end{eqnarray}
 where as in super horizon limit we get an exponentially damped solution of the form
 \begin{eqnarray}
 \label{ega9}
 v_e \propto a^{\dfrac{(\epsilon - 1)(1 + \sqrt{1 + 4\omega^2})}{2}} \ or \quad  a^{\dfrac{(\epsilon - 1)(1 - \sqrt{1 + 4\omega^2})}{2}}.
 \end{eqnarray}
From Eq. (\ref{ega9}) we get,  $v_e \sim a^{-(1+\lambda - \epsilon)}$ or as $ v_e \sim a^{\lambda}$ so that $h_e$  has two solutions  $a^{-(1+2\lambda-\epsilon)}$ and a constant. This is how we expect the gravitational wave to behave once it crosses the horizon; that means the amplitude of oscillation becomes negligible in comparison to the wavelength and the wavelength is said to be frozen.
    
                   The exact solutions can be obtained by rewriting  Eq. (\ref{ega7}) in terms of a  new variable $x = -  k\eta $  followed by a redefinition  $v_e \to \bar{v}_e = x^{-1/2} v_e$. Now Eq. (\ref{ega7}) can be written in  the form of  Bessel's differential equation given by
   \begin{eqnarray}
   \label{ega11}
   x^2 \dfrac{d^2 \bar{v}_e}{dx^2} + x \dfrac{d\bar{v}_e}{dx} + (x^2 - \nu^2) \bar{v}_e
= o,
   \end{eqnarray}
   Where $\nu^2 = \omega^2 + \frac{1}{4}$ and $\nu$ has the explicit expression
   \begin{eqnarray}
  \label{ega12}
  \nu = \frac{3}{2}  +  \epsilon +  \frac{6\xi + 2\zeta}{12\xi + 5\zeta} \delta 
  \end{eqnarray}
     We define  $-k \eta $ as $x$  because the conformal time $\eta$ varies from  $- \infty $ to 0  during inflation and  $x$ always remains positive.  The exact solutions to the Eq. (\ref{ega11}) can be identified as  the two Hankel functions of first and second kind so that  $ v_e $ is obtained as
   \begin{eqnarray}
   \label{ega13}
   v_e = \sqrt{x}\bar{v}_e = \sqrt{- k \eta}  \ \left[ C_1 H_{\nu}^{(1)}(x) + C_2 H_{\nu}^{(2)}(x) \right].
   \end{eqnarray}
     In the asymptotic limit of large $x$ ($x >> 1$)i.e in the deep subhorizon limit  the two Hankel functions takes the approximate form as
     \begin{eqnarray}
     \label{ega14}
     H_{\nu}^{(1)}(x>>1) \sim \sqrt{\dfrac{2}{\pi x}} e^{-\dfrac{i \pi}{4}(1+2\nu )}  e^{ix}\nonumber\\
          H_{\nu}^{(2)}(x>>1) \sim \sqrt{\dfrac{2}{\pi x}} e^{\dfrac{i \pi}{4}(1+2\nu )}  e^{-ix}
     \end{eqnarray}
     The second term in the solution(\ref{ega13}) in the subhorizon limit is the diverging  solution. Therefore  $C_2$ is taken to be zero and  $C_1$  is  obtained by assuming that the solution matches  with the normalized plane wave solution  $e^{-ik \eta}/\sqrt{2 k}$  which is introduced by Eq. (\ref{ega7}) as an initial condition. This hypothesis can be viewed as a direct consequence of Bunch-Davies vacuum condition. So $C_1$ is given by
     \begin{eqnarray}
     \label{ega15}
     C_1 = \dfrac{1}{2} \sqrt{\dfrac{\pi}{k}} e^{i \left(\nu + \dfrac{1}{2}\right)\dfrac{\pi}{2}},
     \end{eqnarray}
     and the final solution for $v_e$ is given by
     \begin{eqnarray}
     \label{ega16}
     v_e(k, \eta) = \dfrac{\pi}{2}  \ e^{i\left(\nu + \dfrac{1}{2}\right)\dfrac{\pi}{2} } \sqrt{-\eta} \  H_{\nu}^{1}(-k\eta).
     \end{eqnarray}
     \subsection{Super horizon modes}
      Once a scale grows beyond the horizon, their amplitude freezes and after the end of inflation these scales re-enter the horizon after the end of inflation during radiation and matter dominated era. Especially the modes  which have grown beyond the horizon at least 60 e-folds before the end of inflation are important as they re-enter the horizon during radiation domination era and lay imprint on the CMB surface. In the superhorizon limit i.e  when  $x$ is smaller then unity, the Hankel function $H_{\nu}^{(1)}(x)$ takes the asymptotic form
     \begin{eqnarray}
     \label{egb1}
     H_{\nu}^{(1)} (x << 1) \sim \dfrac{\Gamma (\nu)}{\pi} e^{i \frac{\pi}{2}} \left( \frac{x}{2}\right)^{-\nu}.
     \end{eqnarray}
     Substituting the asymptotic form of $H_{\nu}^{(1)}(x)$ in Eq. (\ref{ega16}), we get the solution for $v_e$ in super horizon regime as
     \begin{eqnarray}
     \label{egb2}
     v_e(k, \eta) = \dfrac{\Gamma (\nu)}{\sqrt{ k \pi}}  \ 2^{\nu - 1} e^{i\left(\nu  -  \frac{1}{2}\right)\frac{\pi}{2}} (-k\eta)^{\frac{1}{2} - \nu}   . 
      \end{eqnarray}
  In quasi de-Sitter space the scale factor $a$ varies as $\eta^{-(1 + \epsilon)}$.  So the gravitational wave in superhorizon limit takes the form
    \begin{eqnarray}
    \label{egb3}
    h_e = a^{-\lambda} v_e = \dfrac{A}{k^{\lambda + \epsilon + \frac{1}{2}}} (-k\eta)^{\lambda + \epsilon + \frac{1}{2} - \nu},
\end{eqnarray}      
 where  $A = \Gamma (\nu) 2^{\nu - 1} e^{i(\nu - 1/2)\pi /2} / \sqrt{\pi} $ . As $\nu $ is equal to $ \lambda + \epsilon + \frac{1}{2} $, so $h_e$ becomes a constant. But  $h_e $ is not independent of the scales rather $h_e \propto k^{-(\lambda + \epsilon + \frac{1}{2})} \ or \  k^{-\nu}$. In exact de-Sitter type inflation, the r.m.s amplitude $k^{3/2} h_e $ is independent of scales. In quasi de-Sitter scenario, though it depends on $k$  but this dependence is quite small where the exponent of $k$ is a combination of slow roll parameters only. In the general  picture, the amplitude keeps on decreasing as the scales grow beyond the horizon. The power spectrum of the tensor perturbation  is obtained as 
   \begin{eqnarray}
   \label{egb4}
   P_T(k) = \dfrac{k^3}{2\pi^2} \sum_e |h_e|^2 =  \dfrac{|A|^2}{\pi^2} k^{2(1 - \lambda - \epsilon )},
\end{eqnarray}    
and from Eq. (\ref{egb4}) the spectral index $n_T$  can be obtained as 
  \begin{eqnarray}
  \label{egb6}
  n_T = \dfrac{dlnP_T}{dlnk}\Big|_{aH = k} = -2\epsilon -\dfrac{12\xi + 4\zeta}{12 \xi + 5\zeta} \delta .
  \end{eqnarray}
  So the power spectrum in our model is not exactly independent of scales rather it shows nearly scale invariant nature. Similarly from the expression of $n_T$ in Eq. (\ref{egb6}), the difference from a usual scalar field inflation model can be observed to be in the second term that is proportional to the slowroll parameter $\delta$. 

\section{\label{concl}Conclusion}      
As an initial step towards studying the cosmological perturbation theory of inflation model(s) with antisymmetric tensor field $B_{\mu \nu}$, we included tensor perturbations to the background FLRW metric and analyzed their dynamics. For completeness, we generalized the analysis of background dynamics of the model first presented in Ref. \cite{aashish2018,aashish2019} taking into account all nonminimal coupling terms upto linear order in curvature, and found the parameter space region within which stable de-Sitter solutions are expected (Fig. \ref{fig2}).

Secondly we checked the possibility of ghost instability in tensor modes and found the region of allowed parameter space where tensor modes are free from ghost instability (Fig. \ref{fig3}). Further we evaluated the speed of gravitational waves and found appropriate constraints for parameters to match the observed speed of propagation of gravitational waves from the recent gravitational wave data GW170817 which demands the speed of the gravitational wave to be approximately equal to the speed of light. It turns out that the nonminimal coupling with $R_{\mu \nu}$ plays an essential role here since the condition $\tau \approx \zeta / 2\kappa $ fixes the speed of gravitational waves to be unity in natural units. As alluded to before, an interesting aspect that can be explored in future is to check if metric transformations similar to that of Ref. \cite{creminelli2014n} exist for our model, that can result in $c_{T} = 1$. 
         
We also studied the evolution of gravitational waves and derived it's solution in the subhorizon and the superhorizon limit.  The superhorizon mode is found to be constant which only depend on the scales. The dependance of  the rms amplitude of the superhorizon mode  on k is very small i.e. through the slow roll parameters. Finally we calculated the power spectrum and the tensor spectral index for completeness. The spectral index $n_T$ differs from that of a scalar field inflation model with additional term proportional to the slow roll parameter $\delta$ which can help us achieving the requirements of CMB observations. 

We leave the scalar and vector perturbations and the observational prospects of this model to be studied in future. However, we speculate that upon taking into account the constraints due to the inclusion of scalar and vector perturbations from the metric side, the resulting parameter space region will lie within the region identified in this work. This expectation is due to the fact that our current results take into account the constraints from our previous analsyses in Ref. \cite{aashish2019} where scalar and vector modes from antisymmetric tensor have already been taken into account. Similarly, it would be interesting to study second order contributions to gravitational waves where (first order) scalar and vector modes may act as source and provide phenomenologically interesting results \cite{baumann2007n, alabidi2012n, zhang2020n}.

\section{Acknowledgments}
This work is partially supported by DST (Govt. of India) Grant No. SERB/PHY/2017041.

\bibliography{reference3a,reference3b,reference3c}
\end{document}